\documentclass[conference]{IEEEtran}
\IEEEoverridecommandlockouts
\usepackage{cite}
\usepackage{amsmath,amssymb,amsfonts}
\usepackage{algorithmic}
\usepackage{graphicx}
\usepackage{textcomp}
\usepackage{xcolor}
\def\BibTeX{{\rm B\kern-.05em{\sc i\kern-.025em b}\kern-.08em
    T\kern-.1667em\lower.7ex\hbox{E}\kern-.125emX}}
\begin{document}

\title{Visualization - a vital decision driving tool for enterprises\\
{\footnotesize }
\thanks
}

\author{\IEEEauthorblockN{\textsuperscript}{Rajath Chikkatur Srinivasa, Supriya Arun, Lauren James, Ying Zhu}
\IEEEauthorblockA{\textit{Department of Computer Science} \\
\textit{Georgia State University}\\
\{rcs1, ljames26, sarun1,\}@student.gsu.edu, yzhu@gsu.edu}
}

\maketitle

\begin{abstract}
This report documents the results found through surveys and interviews on how visualizations help the employees in their workspace. The objectives of this study were to get in-depth knowledge on what prepares an employee to have the right skill set in constructing an informative visualization as well as the tools and techniques that they use on their daily basis for analysis and visualization purposes. Using the results gathered, we sorted the information in many different ways for analysis and came to conclusions ranging from corporation-based strategies to individualized employee and position preferences.
\end{abstract}

\vspace{20pt}
\section{Introduction}
Since the ease of access to technology has ensured a data explosion, we can say that the 21st century is the era of data. Due to this, industries have started to make their decisions based on the data they receive from the consumers. To make sense of the data that the companies receive, it is crucial to have visualizations for straightforward interpretation since the raw format of the data itself is not understandable for everyone. Besides, visualizations are used for a straightforward interpretation of data and accessible communication and/or explanation between the technical and non-technical sides of the industry. This raised a question on which part of the industry utilizes data visualizations more commonly than the other.

Through this project, we explore the impact of visualizations in the workplace by directly learning from employees how they interact with visualizations in their day-to-day tasks. Our main goal is to discover the challenges and benefits that employees encounter while using them and whether or not they go through a training process to understand better how to build and work with visualizations. The method that we used to gather our data is through surveys and interviews. The data was collected from a variety of positions to get a bigger picture of how visualizations have an impact on the industry. Based on the results collected, we categorize it according to the company and determine which roles have a higher priority in undergoing a training process and which positions have more knowledge on a variety of visualization techniques.

\section{Related Work}
L.A.Shah et al \cite{Shah2018} analyzes performance measurement system’s output by leveraging graph visualizations. They distribute visualizations based on area which helps in making targeted decisions and showcases the importance of visualization to industries.

Sauter et al \cite{Sauter2011} is a case study that studies the development of visualization to represent correlated data. It aims to address specific visualization discussions like choice of visualization technology and visualization algorithms. This helps enterprises to narrow down decisions on visualization.

Alharthi et al \cite{Alharthi2017} applied techniques to visualize Hajj season data. This showcases that visualizations can be a driving factor for deciding various types of events and activities.

Cota et al \cite{Cota2014} describe how using advanced visualization tools within business intelligence software helps drive decision makings within meetings. This work is an example for how enterprise decision making leverages visualizations.

Duru et al’s \cite{Duru2013} work provides a theoretical explanation on what impact software visualization tools creates in software industry and how it’s lack of usage is derailing decision making. The tested it with actual programmers to witness the difference between the outcome of understanding a business use-case with and without visualization.

\section{SECTIONS OF THE SURVEY}

\subsection{First part - Personal experience}

For this study, we decided to interview employees from six companies: Bank of America, State Farm, N2N Services, Verizon, IBM, and Informatics. Each company surveyed had representatives from technical and non-technical positions. The technical positions consisted of Software Developers, Software Engineers, Data Analysts, Data Scientists, Business Analysts, Product Managers, and Product Developers. Non-technical positions included Project Managers, Sales Managers, Operations Team, CEOs, CTOs, COOs, MDs, Founders, Management Team, and HR/Recruiters. 

\begin{figure}[htbp]
\centerline{\includegraphics[width=7.5cm]{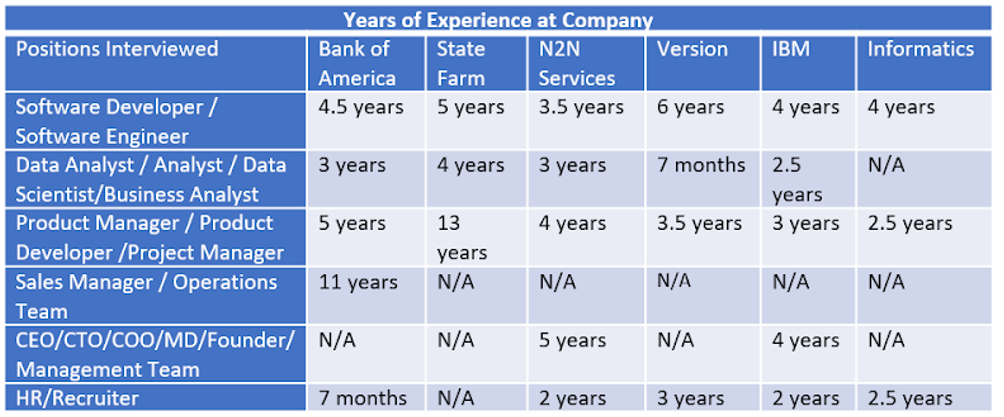}}
\caption{Years of experience at company}
\label{fig}
\end{figure}
The study started by inquiring the employees on their personal experience with making use of visualizations in their workplace. Questions such as how often they use and how they incorporate it in their work were being asked to analyze in which format visualizations are primarily being used. The most popular response for the frequency of usage was “Sometimes,” with it having 61.5 percent of the votes, while the second-highest was “Every day” with 30.8 percent of the votes, and the last one was “Not at all.” 

\begin{figure}[htbp]
\centerline{\includegraphics[width=7.5cm]{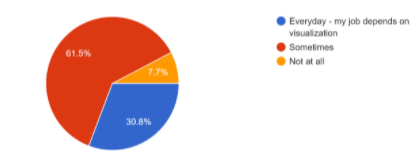}}
\caption{How often do the employees use data visualization in their work?}
\label{fig}
\end{figure}

Just by viewing these results without any background details, it could be said that the employees who voted for “Not at all” are from positions that do not require any form of data to be viewed for decision making, such as recruiters or administrators. On the other hand, employees who voted for “Sometimes” or “Every day” can be predicted to be in positions like business/data analysis or management. The majority of the employees mention that they use visualizations for reviewing and drawing conclusion purposes, while others say for maintaining timelines and explanations. Regardless of the employee’s direct usage of visualizations for their work assignments, all employees surveyed have been exposed to them in the workplace.

Through this, we can conclude that creating visualizations helps in various ways, and the type and responsibilities of the data visualizations utilized in the workspace tend to correspond with the job position. The job positions that use them for reviewing and drawing conclusions could vary mainly by the employees from the business department. Using it for explanations could be workers from the technical department, while people using visualizations for maintaining agendas could be from the executive department

\subsection{Second part - Prior knowledge and Training}
After inquiring about personal experience, employees were then asked about any knowledge they had on visualizations before coming into the workplace and the training processes they may have participated in after joining the company. This was fulfilled through questions such as what courses or tutorials they used to prepare themselves before joining the company; did they take any separate online courses to feel prepared; and lastly, asking to see if the company trains them separately for visualizations and data analysis. From the results gathered, 92.3 percent of the employees say that their company did not train them in visualization and the same percentage of them also say that they did not undergo any training from the company for data analysis.
\begin{figure}[htbp]
\centerline{\includegraphics[width=7.5cm]{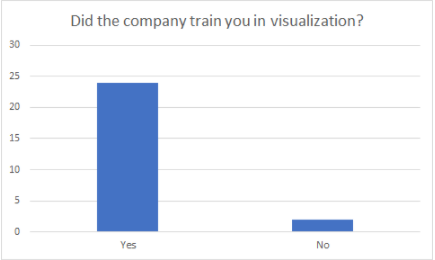}}
\caption{Did the company train you in visualization?}
\label{fig}
\end{figure}

\begin{figure}[htbp]
\centerline{\includegraphics[width=7.5cm]{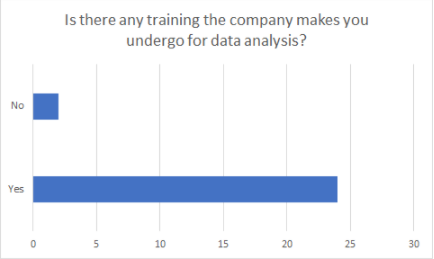}}
\caption{Is there any training the company makes you undergo for data analysis? }
\label{fig}
\end{figure}
Based on this, we can conclude that data analysis and visualizations are topics that the industry already expects the employee to know before joining the company. Employees who answered that they went through a training process said that it was through their company's internal webinar, which had courses that could be completed whenever free. This says that the training process is not one-on-one or hands-on learning; instead, it is just an online course similar to anyone learning data analysis outside the university through online resources. When the topic of prior knowledge on visualizations was asked, 69.2 percent of them said "yes," whereas 30.8 percent reported "no," which could mean that people who answered "yes" already were exposed to this information in their university courses or from their previous workplace. In contrast, people who said "no" would have acquired the knowledge in their current job using visualizations.

\begin{figure}[htbp]
\centerline{\includegraphics[width=7.5cm]{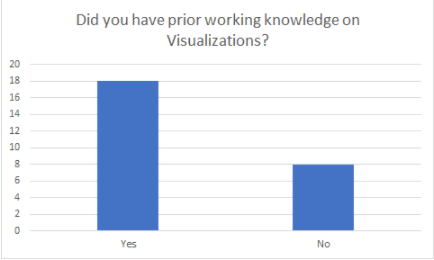}}
\caption{Did you have prior working knowledge in visualizations? }
\label{fig}
\end{figure}

Employees who answered that they did have prior knowledge on the concept of visualizations also mentioned that they obtained the knowledge through courses on Databases and Introduction to BI, and Introduction to Data Science. Some also mention that they learned it in their job itself. The majority concluded that their courses taken in their university were sufficient in preparing them for the visualization knowledge they have currently. Some added that once they got through their first presentation with some visualizations, they got familiar with the concepts and became comfortable with them by their next presentation. 76.9 percent said they did not use any online materials to educate themselves on visualizations, whereas 23.1 percent said they used online resources to succeed in that concept.

\begin{figure}[htbp]
\centerline{\includegraphics[width=7.5cm]{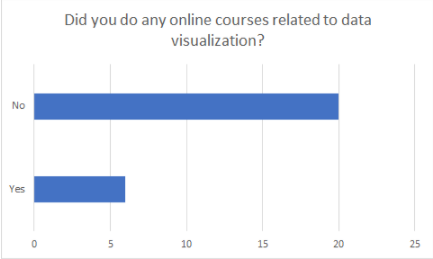}}
\caption{Did you do any online courses related to data visualization? }
\label{fig}
\end{figure}
It can be said that people who did learn it in their job itself tend to be from a department that does not require complex visualizations for decision making and it also shows that building a foundation in visualization is not a highly challenging task. 

\subsection{Third part - Challenges and Complaints} 
	As we continued with the survey questions, we asked the employees 
about the challenges they face and their complaints regarding their visualizations, the ones they come across at work, or even just in general. We then ask them how it can be less challenging to get their perspective on how visualizations can be improved or updated and we also ask what they think is a good visualization. Employees mention that the topmost complaint is that the in-built visualizations are not as flexible in terms of how it suits their data set or how much data can be viewed at once. Some also say that the data can get too cluttered, so it sometimes defeats its purpose. They added that the end-users sometimes do not understand the meaning behind viewing the data in a visualized format. Through these thoughts, it can be concluded that visualizations that are already built-in are for depicting simple visualizations such as bar graphs or line graphs that incorporate simple forms of data. This is why employees might have complained about the built-in ones, but if they had to code the visualizations, they would not have such complaints on flexibility and the amount of data shown since they are manually programming the visualization. For complaints on end-users not clearly understanding the importance of data to be viewed in a visualized manner could be from a background that has to do with quick decision making where they do not have time or the necessity to view the collected information throughout the years or the opinions from the customer. 

The importance of visualizations vary from the time and the importance of the scenario, for example, if the business is taking a huge step in how they function or how they would deal with their stakeholders, then visualizations play a crucial role because it is essential to see all the treads, patterns and details before making a decision. However, if a decision does not require much research before making a move, then a detailed visualization with many comparisons is unnecessary even though it is good. Many employees say that the challenging part of visualizations is to know which type is the suitable one for the data set and which one conveys the right meaning from the results they have found. Some add that explaining visualizations is hard sometimes, and it is due to a lack of understanding from the end-users.

Through this, it can be said that keeping the visualizations as simple as possible can be the key for both cases. Simple visualizations such as bar graphs, pie charts, etc., can be drawn from most of the data sets and these types of depictions are less challenging for explanation and for showcasing the intended meaning behind the visualization made. Understanding the visualizations for the stakeholders and end-users can indeed be challenging, but it depends from person to person because some individuals want to talk business rather than using outside tools to serve their point. When they were asked what could make it less challenging, a large number said that knowing how the team understands better is the key. They also say that familiarizing with the basics for succeeding in visualizations is beneficial for any group of people. These are valid points because it takes time to understand how the team depicts better understanding and it all comes down to being a learning process. They also add that they come across visualizations that are overly done in terms of color formatting and incorporating a vast amount of data into one visualization. This is not so appealing because it makes it less readable and distracts the viewers, which makes them not pay close attention. They also add that some visualizations do not show the entire data set that is being there, which is a problem with accurate conclusions. This comes down to picking the correct type of visualization for accurate display and easy understanding. 

This brings up what makes a good visualization and the popular response was simplicity. This is agreeable because the other party is not seeing how complex a visualization is or their skills in creating one; they want to see the point of the visualization, meaning what they learned from that. They also added that the best way to present a visualization is to explain in a detailed manner. Additionally, they say that keeping visualizations based on only one factor per visualization is the most effective way to get the point across. This is crucial because the more data is shown in a single visualization, the more confusing it gets. Simplicity with the right kind of visualization is the key. 

\subsection{Fourth part - Tools and Techniques} 
The employees were lastly asked what type of tools and software they are familiar with and which ones they incorporate in their workplace. They were asked to pick the tools and techniques from a given list: Google Data Studio, Google Trends, Google Analytics, Google Charts, Kibana, Dash, R Studio, and Azure Data Explorer. The most familiar tool is Google Analytics, while other products of Google (Google Data Studio, Google Trends, and Google Charts) and R Studio are also commonly recognized among the survey respondents. On the contrary, the unknown tools were Dash and Kibana.
\begin{figure}[htbp]
\centerline{\includegraphics[width=8cm]{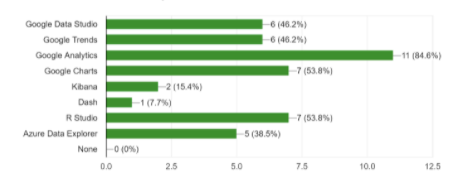}} 
\caption{Which of the below tools have you used or heard before? }
\label{fig}
\end{figure}

As for software applications, there were several selections listed that no employee surveyed had been introduced to for visualizations, including Alteryx, TIBCO SpotFire, ChartBlocks, Datawrapper, Orange, Looker, FineReport, and LeafLet. However, the majority of the employees are familiar with Tableau and Microsoft Excel. Our survey results showed that QlikView, PowerBI, SAS, and Grafana are also tools the company employees have been exposed to. We can conclude that companies may likely provide Microsoft licensing for their employees, and Tableau is a popular data visualization creation tool for people from non-technical and technical backgrounds. 

\begin{figure}[htbp]
\centerline{\includegraphics[width=8.5cm]{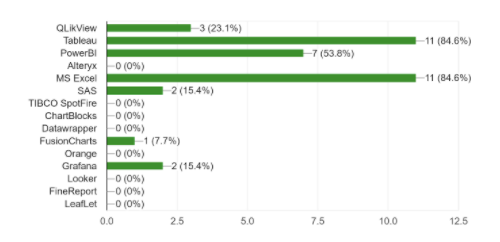}} 
\caption{Which of the following software have you used and are familiar with for visualizations? }
\label{fig}
\end{figure}

We asked a technical question about their exposure to libraries used to code visualizations to continue our in-depth research on the individual’s experience with data visualization tools and applications. The employees from non-technical positions were not familiar with the libraries. Plotly was the most familiar library amongst the technical employees, followed by ggplot, seaborn, and matplotlib. This implies that the technical employees tend to prefer using Python or R when coding visualizations.  

\begin{figure}[htbp]
\centerline{\includegraphics[width=8.5cm]{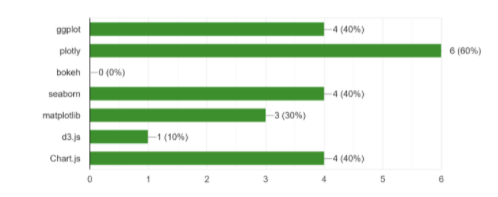}} 
\caption{Which of the following data visualization libraries have you previously used or heard about?  }
\label{fig}
\end{figure}
This survey exposed that basic knowledge of data visualizations is fundamental in any field. We asked our survey respondents about the plotting techniques they are aware of and at least one person was familiar with each type of plot listed. There was only one individual, a Product Manager, who reported that they were unfamiliar with any techniques. The most recognized plot types are scatterplots, bar plots, heatmaps, line plots, and cluster maps. This is a direct result of the most basic visualizations commonly displayed in the workplace, online training, and even university settings.

\begin{figure}[htbp]
\centerline{\includegraphics[width=8.5cm]{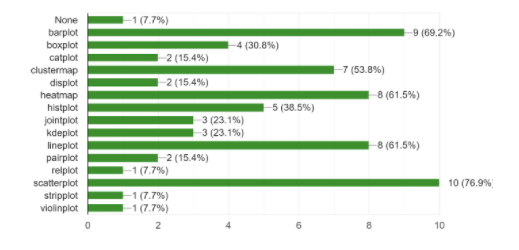}} 
\caption{Which of the below types of plotting techniques have you heard of before?  }
\label{fig}
\end{figure}
The most common domain that the employees have noticed in data visualizations is Information Technology. This was an expected outcome as technical positions tend to use data visualizations more often than non-technical job assignments. The non-technical employees may have been exposed to visualizations when viewing these technical presentations.
\begin{figure}[htbp]
\centerline{\includegraphics[width=8.5cm]{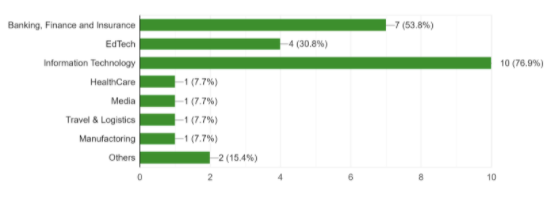}} 
\caption{For which domain have you applied data visualization or observed being applied? }
\label{fig}
\end{figure}

We questioned the industry professionals on their confidence with specific concepts related to visualizations. The professionals were most confident with Exploratory Data Analysis and were significantly less confident with the remaining topics: Time Series, World Cloud, and Kernel Density Estimator. However, there are some outliers as the same Product Manager, unfamiliar with the plotting techniques, was very confident with the Time Series topic. The results made it very noticeable that non-technical industry positions were highly unfamiliar with these concepts, with the overall most unfamiliar topic being the Kernel Density Estimator.

\begin{figure}[htbp]
\centerline{\includegraphics[width=8.5cm]{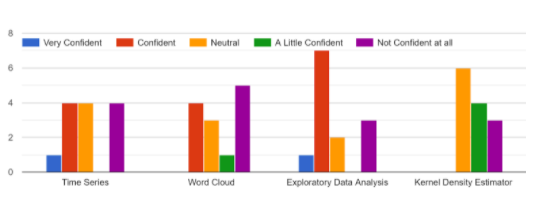}} 
\caption{How confident are you in below topics? }
\label{fig}
\end{figure}

We concluded the survey asking about the impact of data visualizations. All employees agreed that visualizations do not make matters more difficult, but they either maintain or make tasks easier. In terms of communication, understanding, and decision making, a large majority of the individuals surveyed agree that data visualizations are beneficial for people.

\begin{figure}[htbp]
\centerline{\includegraphics[width=8.5cm]{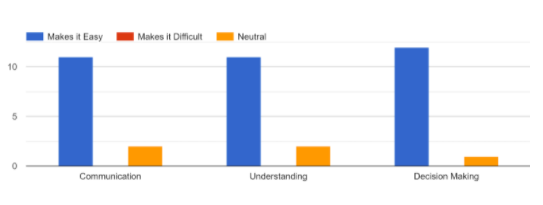}} 
\caption{According to you, what is visualization impact on the below parameter? }
\label{fig}
\end{figure}

It is a proven fact that visualizations communicate data better than words alone as it makes information easier to process. They allow people to comprehend data at faster and more effective rates. Visual representation allows for clearly defined parameters, leaving less room for confusion and miscommunication that leads to wrong decisions. Analyzers can gain insight that they may not have acknowledged without the visualization that encourages rightful decision-making. Large data sets can be easily comprehended by the human brain when it is formatted in a clean, efficient visualization. These visuals make it easier to detect patterns, trends, and outliers in groups of data. Also, data visualizations can answer business questions that are not necessarily intuitively answered. They provide context to make critical data-driven decisions for the financial benefit of an organization. For these reasons, we can assume that these employees value utilizing visualizations in the workplace and would prefer continued usage of these methods.

\section{Results}
The results of this survey were not unanticipated. The most shocking factor to the team was that there was a more extensive representation than expected of employees (7.7 percent) who do not use visualizations in the workplace. We assumed in the present day that every job position should be utilizing data visualizations. We were partially correct in this assumption as each employee revealed they had seen visualizations in work meetings and presentations, but each employee does not need to use them for their job. The most important takeaway from our research is the knowledge we have gained about the employees' perspective. We have learned what employees value when it comes to visualization preferences and learning new information and the patterns in technical versus non-technical positions. The results did not vary based on the company but by position. Our results have also allowed us to generate visualizations about the information companies prioritize exposing their employees through training and presentations.

\section*{Acknowledgment}

Faiz Ahamed Iqbal,Sushmita Suresh,Sneha Krishnan,Vijay Prasad,Vaishnav B,Kiran Kodithala,Bhaskar D,Nicolas E. Colbert,Rakesh Manda,Kaléa Dais,Sumanth Katari,Rathnakara Achary,Joji Thomas,Suyash Kotia,Deepthi Saineni

\bibliographystyle{plain}
\bibliography{bibliography.bib}

\end{document}